\documentclass{aa}  

\usepackage{graphicx}
\usepackage{txfonts}
\newcommand\gaia{\textit{Gaia}}

\begin{document}

   \title{Classical Cepheid period-Wesenheit-metallicity relation in the \gaia\ bands.\thanks{Table 1 is only available in electronic form
at the CDS via an anonymous ftp to cdsarc.u-strasbg.fr (130.79.128.5)
or via http://cdsweb.u-strasbg.fr/cgi-bin/qcat?J/A+A/}}


   \author{V. Ripepi
          \inst{1}
          \and
          G. Catanzaro \inst{2}
          \and 
          G. Clementini \inst{3}
          \and 
          G. De Somma \inst{1,4}
          \and    
          R. Drimmel \inst{5}
          \and
          S. Leccia \inst{1}
          \and
          M. Marconi \inst{1} 
          \and
          R. Molinaro \inst{1}
          \and 
          I. Musella \inst{1}
          \and
          E. Poggio\inst{6,5}
          }

\institute{ INAF-Osservatorio Astronomico di Capodimonte, Salita Moiariello 16, 80131, Naples, Italy\\  \email{vincenzo.ripepi@inaf.it}
\and
INAF-Osservatorio Astrofisico di Catania, Via S.Sofia 78, 95123, Catania, Italy 
             \and
INAF-Osservatorio di Astrofisica e Scienza dello Spazio, Via Gobetti 93/3, I-40129 Bologna, Italy 
             \and
Istituto Nazionale di Fisica Nucleare (INFN)-Sez. di Napoli, Via Cinthia, 80126 Napoli, Italy     \and        
Osservatorio Astrofisico di Torino, Istituto Nazionale di Astrofisica (INAF), I-10025 Pino Torinese, Italy
             \and
Universit\'e C\^ote d’Azur, Observatoire de la C\^ote d’Azur, CNRS, Laboratoire Lagrange, France
             }

   \date{}

 
  \abstract
   {Classical Cepheids (DCEPs) represent a fundamental tool to calibrate the extragalactic distance scale. However, they are also
   powerful stellar population tracers 
   in the context of Galactic studies. 
   The 
   forthcoming  Data Release 3 (DR3) of the \gaia\ mission will allow us to study, with unprecedented detail, the structure, the dynamics, and the chemical properties of the Galactic disc, and in particular of the spiral arms, where most Galactic DCEPs reside.}
   {In this paper we aim to quantify the metallicity dependence of the Galactic DCEPs' period-Wesenheit ($PWZ$) relation in the \gaia\ bands.}
   {We adopted a sample of 499 DCEPs with metal abundances  from  high-resolution spectroscopy, in conjunction with Gaia Early Data Release 3 parallaxes and photometry to calibrate a $PWZ$ relation in the \gaia\ bands.}
   {We find a significant metallicity term, of the order of $-$0.5 mag/dex, which is larger than 
   the values measured in the near-infrared (NIR) bands by different authors. Our best $PWZ$ relation is $W=(-5.988\pm0.018)-(3.176\pm0.044)(\log P-1.0)-(0.520\pm0.090){\rm [Fe/H]}$.
We validated our $PWZ$ relations by using the distance to the Large Magellanic Cloud as a benchmark, finding very good agreement with the geometric distance provided by eclipsing binaries. 
As an additional test, we evaluated the metallicity gradient of the young Galactic disc, finding $-0.0527 \pm 0.0022$ dex/kpc, which is in very good agreement with previous results.}
   {}


   \maketitle
%

\section{Introduction}

Since their discovery, the period-luminosity (PL) and period-Wesenheit (PW) relations for classical Cepheids (DCEPs) represent the fundamental tools at the basis of the extra-galactic distance ladder  \citep[e.g.][]{Leavitt1912,Madore1982,Caputo2000,Riess2016}. However, the DCEPs are also important astrophysical objects in the context of Galactic studies. Indeed, given their young age ($\sim$50-500 Mys), they are preferentially located in the thin disc of the Milky Way (MW). In particular, thanks to the precise distances that can be derived from the above-mentioned relations, DCEPs can be used to model the disc and, given their young age, to trace their birthplaces in the spiral arms, where star formation is more active. 
In this context, \citet[][]{Chen2019} used more than 1300 DCEPs to model the stellar disc, finding that it follows the gas disc and extends to at least 20 kpc. They also found that the line of nodes of the Galactic disc warp is not oriented in the Galactic centre--Sun direction. Similarly, \citet{Skowron2019}, based on the positions and distances of more than 2600 DCEPs, built a three-dimensional map of the MW, showing the structure of the MW's young stellar population and constraining the warped shape of the MW's disc and proposed a simple model of star formation in the spiral arms. More recently, \citet{Poggio2021} adopted a sample of about 2900 DCEPs, together with open cluster and upper main sequence stars to map the density variations in the distribution of these objects. They found that the DCEP over-densities likely extend the spiral arm portion on a larger scale, that is $\sim$10 kpc from the Sun.
In addition to these studies, when the chemical abundance of the DCEPs is known, they can be used to trace the metallicity gradient of the Galaxy, as was done by \citet[][and references therein]{Genovali2014}, who, for example, found a linear gradient over a broad range of Galactocentric distances between 5 and 19 kpc. 
This result was also later confirmed by \citet{Luck2018} on the basis of homogeneous chemical abundances and {\it Gaia} Data Release 2 parallaxes \citep{Gaia2016,Gaia2018}.

In this context, a great advance is expected by the 
publication of Data Release 3 (DR3) of the {\it Gaia} mission. This release will include astrophysical parameters, such as effective temperature, gravity, metallicity, and extinction, for more than one billion stars which will complement the astrometry and photometry already published in Early Data Release 3 \citep[EDR3][]{Gaia2021}. These unique datasets will allow us to study the structure, kinematics, and chemo-dynamical properties of the Galactic disc with unprecedented accuracy. However, to fully exploit this information,  we need precise distances up to the limits of the Galactic 
disc, for example, at more than 20 kpc from the 
MW disc or 12-15 kpc from the Sun. In such distant portions of the Galaxy, even though {\it Gaia} photometry and proper motions remain sufficiently precise,  parallaxes will not be able to provide distances with the precision required to
provide an accurate mapping of the positions and kinematics of the disc 
 at the level of 5-10\%. The DCEP $PL$ and $PW$ relations can supply distances at the required precision; however, 
it is 
 crucial to have these relations calibrated in the {\it Gaia} bands in order to exploit the exquisite photometry provided by {\it Gaia} and to incorporate a metallicity term. Indeed, even though it has been known for a long time that the DCEP $PL$ and $PW$ relations should depend on metallicity \citep[see e.g.][and references therein]{Caputo2000,Fiorentino2002,Marconi2005,Romaniello2008,Bono2010,Freedman2011,Riess2016}, it was only the advent of {\it Gaia} that allowed us to make a more precise estimate about the size of such a dependence. The  period-luminosity-metallicity ($PLZ$) and period-Wesenheit-metallicity ($PWZ$) relations in the near-infrared (NIR) bands based on DR2 parallaxes 
  provided inconclusive results \citep[][]{Groenewegen2018,Ripepi2020}, owing to the still insufficient precision of the DR2 astrometry. \citet{Ripepi2019} used similar data to calculate the first $PLZ$/$PWZ$ relations in the {\it Gaia} bands, obtaining again partially significant metallicity terms. The publication of EDR3 improved the situation significantly and, for example, \citet{Riess2021} and \citet[][R21 hereafter]{Ripepi2021} obtained significant $PLZ$/$PWZ$ relations in a variety of optical and NIR filters. As for the {\it Gaia} bands, in a previous work \citep{Poggio2021}, we adopted a sample of 852 fundamental-mode (F-mode)  and  396 first overtone (1O-mode)  DCEPs with usable EDR3 parallaxes and a confirmed classification  to calibrate different $PW$ relations in the {\it Gaia} bands for the two pulsation modes, but not including the metallicity term as this information was missing for most of the calibrating DCEPs. 
  
The purpose of this paper is 
to include the dependence on metallicity and calculate the $PWZ$ relations in the {\it Gaia} bands.
This will allow us to exploit the data products of DR3, which will include  individual metallicities from the Radial Velocity Spectrometer \citep[RVS][]{Gaia2016} for a consistent sample of Galactic DCEPs (e.g. about 1000 objects) with a precision of the order of 0.1 dex \citep[see][and references therein]{Gaia2016}, and, in turn, to obtain the 5\% accurate distances needed for a precise mapping and kinematic study of the MW disc. 

\begin{sidewaystable*}
\caption{Data used in this paper. Only the first ten lines of the table are shown here to guide the reader to its content. The machine readable version of the table will be published at the CDS (Centre de Données astronomiques de Strasbourg, https://cds.u-strasbg.fr/).}
\label{table:data}    
\footnotesize\setlength{\tabcolsep}{5pt}
\centering          
\begin{tabular}{c c c c c c c c c c c c c c c }     
\hline\hline       
  GaiaEDR3\_sourceid     &  Name                             &          RA     &       Dec     &        Mode  &      Period  &     $G$    &   $G_{BP}-G_{RP}$ &    plx   &  plx\_err  & plx\_corr &  RUWE   & ${\rm [Fe/H]}$  & ${\rm [Fe/H]}_{err}$ & Source \\
                        &                                   &         (Deg)   &       (Deg)   &              &        (d)   &    (mag) &    (mag)  &  (mas)   &   (mas)   &  (mas)  &          &  (dex)  & (dex)  &      \\
(1) & (2) &  (3) & (4) & (5) & (6) & (7) & (8) & (9) & (10) & (11) & (12) & (13) \\
\hline
  3430067092837622272   &  AA Gem                           &      91.645608  &    26.329220  &       DCEP\_F &   11.301566  &   9.393  &   1.363  &   0.2749  &   0.0177  &   0.3114 &  1.249  &  -0.08  &   0.12  &  G18 \\
  3102535635624415872   &  AA Mon                           &     104.349041  &    -3.843336  &       DCEP\_F &    3.938148  &  12.185  &   1.869  &   0.3139  &   0.0149  &   0.3163 &  1.163  &  -0.12  &   0.12  &  G18 \\
  4260210878780635904   &  AA Ser                           &     280.340671  &    -1.111234  &       DCEP\_F &   17.142112  &  11.082  &   2.817  &   0.2408  &   0.0301  &   0.2787 &  0.952  &   0.38  &   0.12  &  G18 \\
  473239154746762112    &  AB Cam                           &      56.534386  &    58.784221  &       DCEP\_F &    5.787580  &  11.554  &   1.663  &   0.2128  &   0.0208  &   0.2406 &  1.276  &  -0.11  &   0.12  &  G18 \\
  462252662762965120    &  AC Cam                           &      50.949512  &    59.355669  &       DCEP\_F &    4.156769  &  11.855  &   2.027  &   0.3206  &   0.0178  &   0.3431 &  1.206  &  -0.16  &   0.12  &  G18 \\
  3050050207554658048   &  AC Mon                           &     105.249240  &    -8.708983  &       DCEP\_F &    8.014931  &   9.646  &   1.595  &   0.3549  &   0.0187  &   0.3829 &  1.379  &  -0.06  &   0.12  &  G18 \\
  462407693902385792    &  AD Cam                           &      52.358206  &    60.446467  &       DCEP\_F &   11.263048  &  11.926  &   2.195  &   0.2965  &   0.0174  &   0.3150 &  1.185  &  -0.28  &   0.12  &  G18 \\
  6057514092119497472   &  AD Cru                           &     183.248564  &   -62.096823  &       DCEP\_F &    6.397233  &  10.570  &   1.831  &   0.2929  &   0.0132  &   0.3153 &  1.017  &   0.08  &   0.12  &  G18 \\
  3378049163365268608   &  AD Gem                           &     100.781296  &    20.939106  &       DCEP\_F &    3.787998  &   9.709  &   0.986  &   0.3356  &   0.0197  &   0.3698 &  0.969  &  -0.14  &   0.12  &  G18 \\
  5614312705966204288   &  AD Pup                           &     117.016035  &   -25.577771  &       DCEP\_F &   13.596814  &   9.635  &   1.447  &   0.2331  &   0.0165  &   0.2537 &  1.362  &  -0.06  &   0.12  &  G18 \\
\hline                  
\end{tabular}
\tablefoot{The meaning of the different columns is as follows: (1) \gaia\ EDR3 identification; (2) other name of the DCEP; (3) and (4) equatorial coordinates (J2000); (5) mode of pulsation -- F, 1O, F/1O, and 1O/2O indicate the fundamental, first overtone, and the mixed mode pulsation modes, respectively; and (6) period of pulsation. For mixed mode DCEPs, the longest period is listed; (7) and (8) $G$ magnitude and $G_{BP}-G_{RP}$ colour in the \gaia\ bands, respectively. These quantities are listed without errors as we assumed a conservative uncertainty of 0.02 mag for each \gaia\ band; (9) and (10) original parallax value and error from \gaia\ EDR3 catalogue; (11) parallax value corrected according to \citet{Lindegren2021}; (12) RUWE value from \gaia\ EDR3; (13) and (14) iron abundance and error from literature; and (15) literature source of the iron abundance -- G14=\citet{Genovali2014}, GC17=\citet{Gaia2017}, G18=\citet{Groenewegen2018}, and R21=\citet{Ripepi2021}.}
\end{sidewaystable*}

\section{Adopted sample}  

\begin{figure}
   \centering
   \includegraphics[angle=0,width=8.5cm]{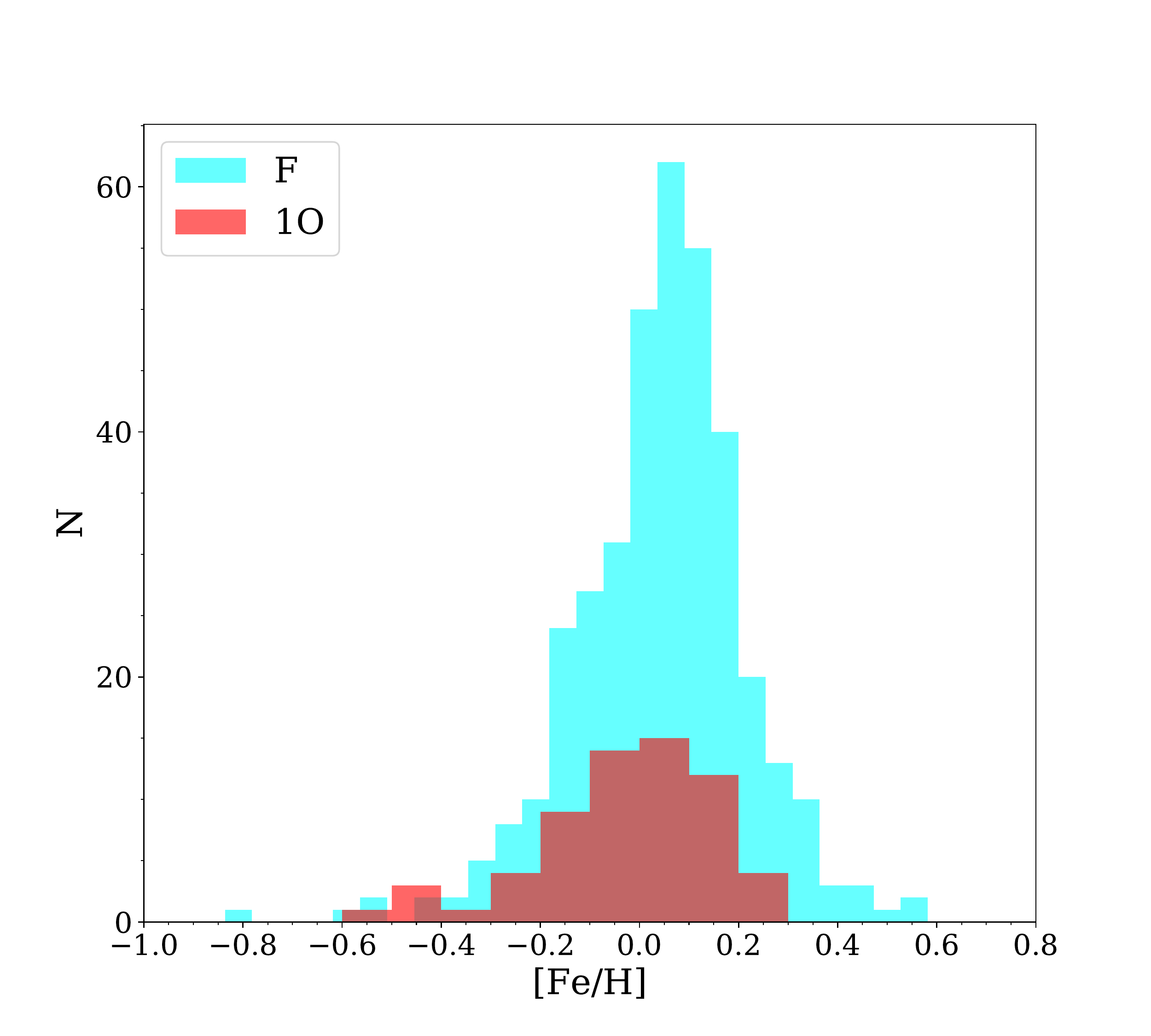}
   \caption{Histogram of the [Fe/H] values of the sample of F and 1O mode DCEPs adopted in this work.}
              \label{fig:histoMet}%
    \end{figure}

To calibrate the $PWZ$ relation in the {\it Gaia} bands, we need a significant sample of DCEPs with a metallicity from high-resolution spectroscopy. We decided to adopt the sample of DCEPs as in R21, which includes 409 F, 68 1O, 18 F/1O, and 4 1O/2O  pulsators. For the mixed-mode Cepheids, we used the longest period of pulsation. The metallicity of DCEPs in our sample were taken from  \citet{Genovali2014}, \citet{Gaia2017}, \citet{Groenewegen2018}, and \citet{Ripepi2021}, and a histogram of their distribution is shown in Fig.~\ref{fig:histoMet}. 

The position of our sample stars was cross-matched with the EDR3 catalogue to retrieve the   $G,G_{BP},G_{RP}$ magnitudes, the parallax  with relative error, and the re-normalised unit weight error (RUWE)\footnote{Section 14.1.2 of 'Gaia Data Release 2 Documentation release 1.2'; https://gea.esac.esa.int/archive/documentation/GDR2/} for  each Cepheid in the sample. The parallax zero point offset (ZPO) was  corrected on an individual basis according to \citet{Lindegren2021} (see R21 for details on the procedure). To maintain the consistency with R21, here we also adopted the global parallax ZPO correction of $-14\pm$6 $\mu$as calculated by \citet[][]{Riess2021} (see R21 for a discussion on this subject).

To ensure that sources  with poor astrometry were not included, we 
retained only  DCEPs with RUWE$<$1.4 and $G>6$ mag (see R21 and references therein). The resulting sample is composed of 372 F- and 63 1O-mode DCEPs. Given the 
limited number of 1O-mode DCEPs in the sample, 
we 
 fundamentalised 
their periods, according to 
the \citet{Feast1997} equation 
$P_F = P_{1O}/(0.716-0.027 \log_{10} P_{1O})$, where $P_F$ and $P_{1O}$ are the F and 1O mode DCEP periods, respectively.
We then fitted F-mode and fundametalised 1O-mode DCEPs all together.

\begin{figure}
   \centering
   \includegraphics[angle=0,width=8.5cm]{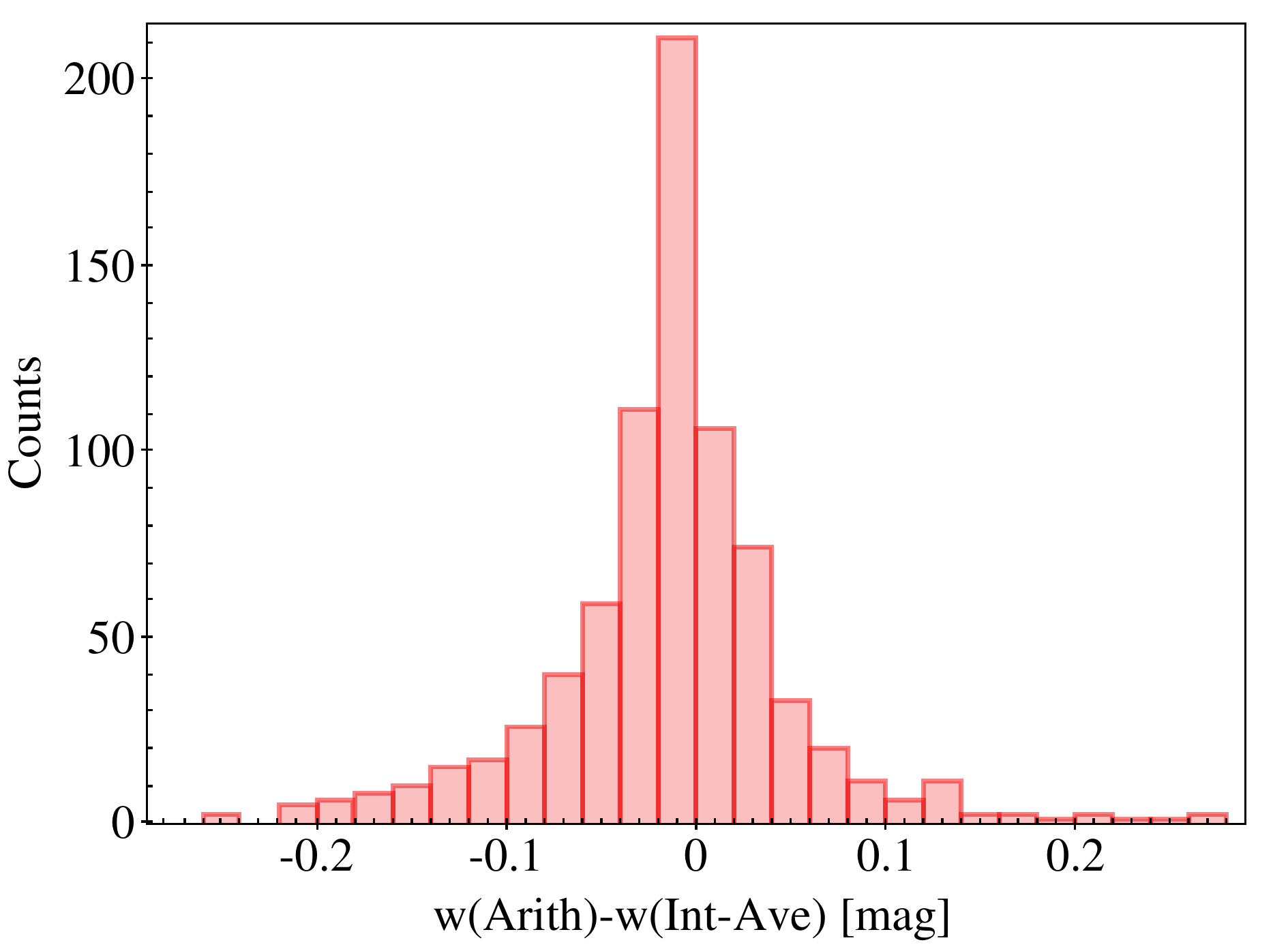}
   \caption{Difference between the Wesenheit magnitude for a sample of DCEPs in Gaia DR2 averaged arithmetically and in intensity from the fit of the light curve.}
              \label{fig:w_diff}%
    \end{figure}

\begin{figure}
   \centering
   \includegraphics[angle=0,width=9.0cm]{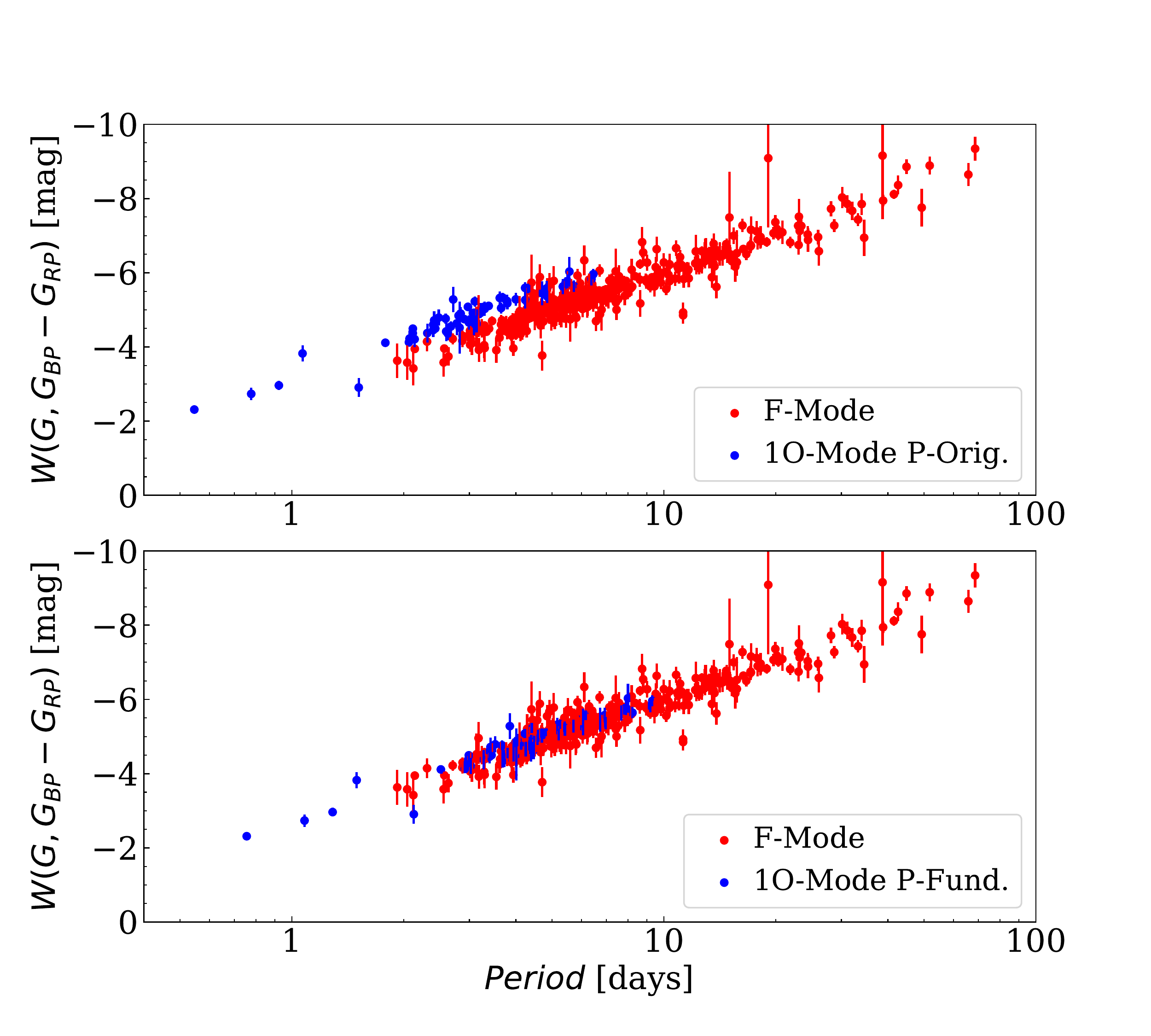}
   \caption{PW relation in the Gaia bands for the programme stars. Red and blue dots represent F- and 1O-mode pulsators, respectively. The top and bottom panels show the PW relation including not fundamentalised and fundamentalised 1O mode DCEPs, respectively.}
              \label{fig:pwGaia}%
    \end{figure}

It is important to note that the correct average magnitude of a DCEP is obtained by modelling the observed light curve with a truncated Fourier series (or other functional forms), integrating the model in intensity and then transforming the result back into magnitude. Since magnitudes in the \gaia\ EDR3 
catalogue are obtained by a simple arithmetic average, they can differ by several hundredths of magnitude from the intensity-averaged magnitudes  \citep[see e.g.][]{Caputo1999}. However, as 
shown in \citet{Poggio2021}, this drawback is greatly mitigated by 
adopting the so-called Wesenheit magnitude ($w$)\footnote{The Wesenheit magnitudes are reddening-free by definition, provided that the extinction law is known \citep[][]{Madore1982}}. In the \gaia\ bands, the coefficient of the $w$ magnitude was derived  empirically by \citet[][]{Ripepi2019} on the basis of the DCEPs in the Large Magellanic Cloud (LMC) as $w=G-1.90 \times (G_{BP}-G_{RP})$. \citet{Poggio2021} found that due to a 
favourable combination of magnitude and colour, the difference between arithmetic and intensity-weighted magnitude is, on average, less than 2\% for 80\% of the DCEPs included in DR2. Here we further investigated 
this issue using 
900 DCEPs reclassified by \citet{Ripepi2019} 
for which both  arithmetic and intensity-averaged magnitudes are available in the \gaia\ DR2 catalogue. The results are shown in Fig.~\ref{fig:w_diff}. Quantitatively, we find a median difference w(Arith)-w(Int-Ave)=-0.01$\pm$0.03 mag. In the following, we thus use arithmetic Wesenheit magnitudes after summing 0.01 mag to their values. Our sample is now ready for the following analysis. Its appearance in the $PW$ plane is shown in Fig.~\ref{fig:pwGaia}.

\begin{table*}
\caption{Results of the determination of the $PWZ$ relation from the fit to the observations. }
\label{tab:results} 
\footnotesize\setlength{\tabcolsep}{3pt}
\centering          
\begin{tabular}{ccccccccccc} 
\hline\hline             
Case & $\alpha$ & $\beta$ & $\gamma$ & n.MW & $\sigma$ & $\chi^2$ & Mode & $\mu_{LMC}$ & n.LMC & Method\\
\hline
1 & $-6.023 \pm 0.014$ &  $-3.301 \pm 0.048$ & 0.0                 &  372 & 0.012 & 1.74 & F & 18.687$\pm$0.024       & 2557 & PhotPar\\
2 & $-6.028 \pm 0.013$ &  $-3.289 \pm 0.039$ & 0.0                 &  435 & 0.010 & 1.65 & F+1O & 18.697$\pm$0.024       & 4634 & PhotPar\\
3 & $-5.948 \pm 0.018$ &  $-3.165 \pm 0.054$ & $-0.725\pm0.098$    &  372 & 0.011 & 1.29 & F & 18.370$\pm$0.049  &  2557 & PhotPar\\         
4 & $-5.965 \pm 0.018$ &  $-3.161 \pm 0.051$ &  $-0.598\pm0.094$   &  435 & 0.009 & 1.29 & F+1O & 18.457$\pm$0.052 & 4634 & PhotPar\\ 
5 & $-6.042 \pm 0.013$ &  $-3.294 \pm 0.049$ &  0.0                &  372 & 0.017 & 2.60 & F & 18.708$\pm$0.024 & 2557 & ABL\\
6 & $-6.047 \pm 0.014$ &  $-3.287 \pm 0.050$ &  0.0                &  435 & 0.015 & 2.49 & F+1O & 18.718$\pm$0.024 & 4634 & ABL\\
7 & $-5.971 \pm 0.017$ &  $-3.178 \pm 0.048$ &  $-0.661 \pm 0.077$ &  372 & 0.016 & 2.29 & F & 18.414$\pm$0.048 & 2557 & ABL\\
8 & $-5.988 \pm 0.018$ &  $-3.176 \pm 0.044$ &  $-0.520 \pm 0.090$ &  435 & 0.014 & 2.26 & F+1O & 18.503$\pm$0.046 & 4634 & ABL\\ 
\hline                                   
\end{tabular}
\tablefoot{The quantities $\alpha$, $\beta$, and $\gamma$ are the coefficient of the $PWZ$ relation described in the text; n.MW is the number of MW DCEPs adopted in each minimisation; $\sigma$ is the standard deviation of the mean of the difference $W-W_{calc}$, where $W$ is the observed absolute Wesenheit magnitude and $W_{calc}$ is that calculated from the coefficients $\alpha$, $\beta$, and $\gamma$ (when applicable) reported in the table; $\chi^2$ reports the reduced value of the $\chi^2$ from its minimisation; Mode identifies the adopted sample; $\mu_{LMC}$ represents the distance modulus of the LMC obtained with the specific $PWZ$ relationship; n.LMC is the number of LMC DCEPs adopted to calculate the $\mu_{LMC}$ value;  and Method identifies the two different techniques adopted to fit the data, with PhotPar indicating the results of the minimisation of Eq.~\ref{eq:chi} and ABL indiciating the results from the minimisation of Eq.~\ref{eqABLZ}. Lines 1--2 and 5--6 report the results for the case in which the metallicity term of the $PW$ relation is null.}
\end{table*}

\section{Analysis}
 
To derive the $PWZ$ relation in the \gaia\ bands, we follow the same approach as in \citet{Poggio2021}, which, in turn, is based on the work by \citet{Riess2021}. 
 
We first define the photometric parallax (in mas) as follows: 

\begin{equation}
\varpi_{phot}=10^{-0.2(w-W-10)}\, ,
\end{equation}

\noindent where $w$ is the apparent Wesenheit magnitude (defined above), while $W$ is the absolute Wesenheit magnitude, which can be written as

\begin{equation}
W=\alpha+\beta(\log_{10} P-1.0)+\gamma {\rm {\rm [Fe/H]}}. 
\label{eq:PWZ}
\end{equation}

Indicating the zero-point corrected parallax from EDR3 with $\varpi_{EDR3}$, we minimise the following quantity:

\begin{equation}
\chi^2=\sum \frac{(\varpi_{EDR3}-\varpi_{phot})^2}{\sigma^2}.    
\label{eq:chi}
\end{equation}
 
\noindent 
Here $\sigma$ is the total error obtained by summing up in quadrature the uncertainty on $\varpi_{EDR3}$ and $\varpi_{phot}$: $\sigma=\sqrt{\sigma_{\varpi_{EDR3}}^2+\sigma_{\varpi_{phot}}^2}$. In addition,  $\sigma_{\varpi_{EDR3}}$ is made of three contributions: the standard error of the parallax as reported in the  EDR3 catalogue, which we conservatively increased by 10\%; the uncertainty on the individual ZPO corrections, that is 13 $\mu$as \citep[][]{Lindegren2021}; and  the uncertainty on the global parallax correction,  which is equal to 6 $\mu$as according to \citet{Riess2021}. 
The uncertainty on the photometric parallax is more tricky to calculate. Considering the equivalence $\delta \varpi/\varpi=\delta D/D$, where $D$ is the distance and the definition of the distance modulus $\mu$=$-5+5\log_{10} D$, after propagating the errors and some algebra we have: $\sigma_{\varpi_{phot}}=0.46 \times \sigma_\mu \times \varpi_{phot}$, where $\sigma_\mu=\sqrt{\sigma_{w}^2+\sigma_{W}^2}$. While  $\sigma_{w}$ is easy to calculate by propagating a conservative error of 0.02 mag in each of the three \gaia\ bands ($G,G_{BP},G_{RP}$), $\sigma_{W}$ is more complex because we need to know the intrinsic dispersion of the relation in advance. 
\citet{Desomma2020} published theoretical $PW$ relations for Cepheids  
in the \gaia\ bands.  
In Table 12 of their paper,  they provide  
intrinsic dispersions of the $PW$ relation 
of the order of 0.06-0.08 mag, depending on the model characteristics. 
We have thus adopted a conservative dispersion of 0.1 mag\footnote{It is worth noticing that the typical dispersion in the NIR bands is smaller, that is $\sim$0.07 mag \citep[][]{Riess2019}.}. As the theoretical $PW$ relation did not include a metallicity term, we added, in quadrature, to this dispersion, the uncertainty in metallicity, using iteratively the coefficient we derived from the minimisation procedure. The procedure converged after a few iterations. 

\begin{figure}
   \centering
   \hbox{
      \includegraphics[angle=0,width=8.7cm]{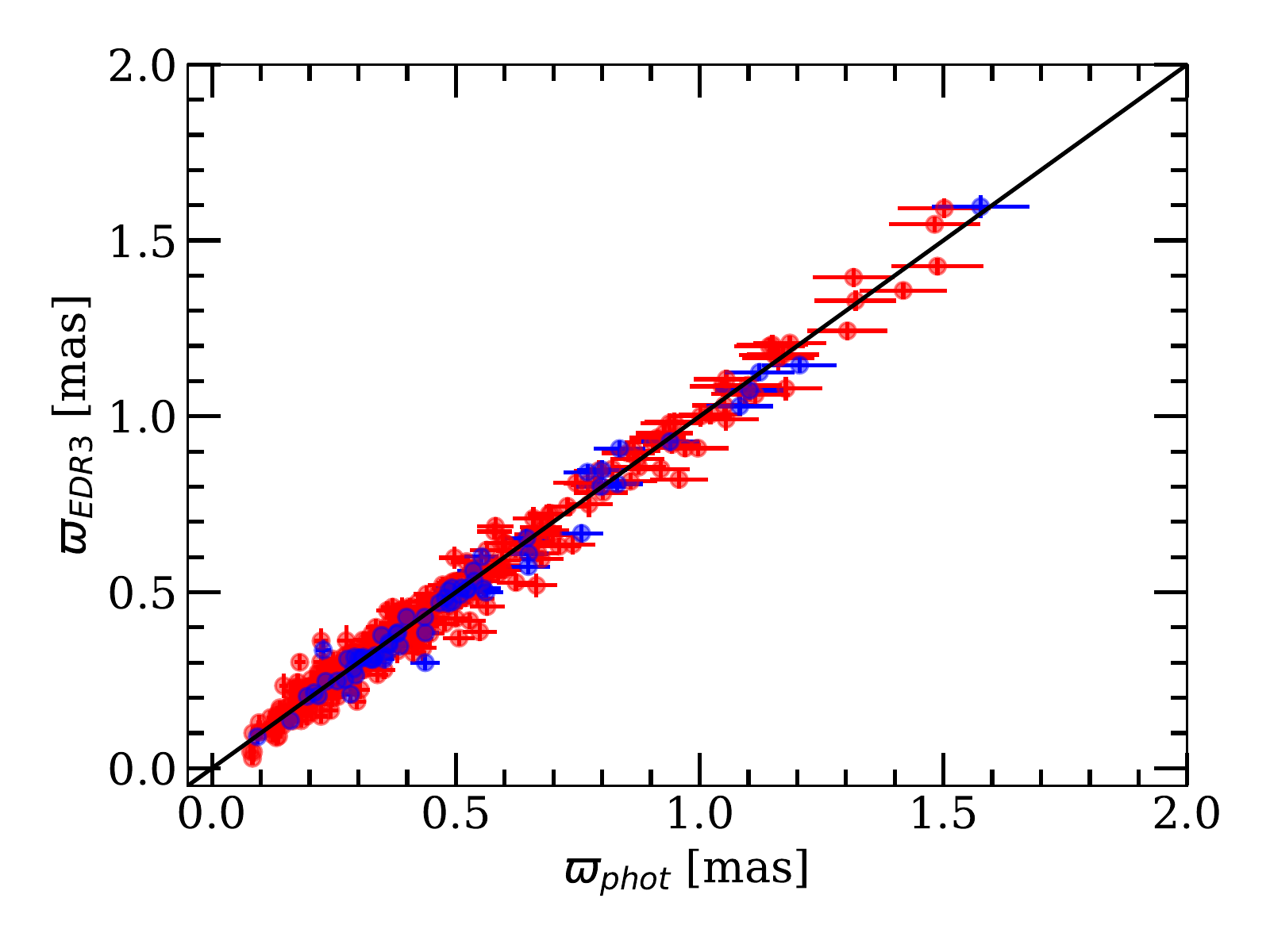}
   }
   \caption{Comparison between the photometric and observed parallaxes for the complete sample (F+1O) of DCEPs. Colours are the same as in Fig.~\ref{fig:pwGaia}.}
              \label{fig:parallaxComparison}%
    \end{figure}

To minimise Eq.~\ref{eq:chi}, we adopted the {\tt python} minimisation routine {\tt optimize.minimize} included in the {\tt Scipy} package \citep[][]{Virtanen2020}. For completeness, we also considered the case in which the metallicity term in the formulation of $W$ is null ($\gamma$=0). The results of the procedure in the case of only F- and of F+1O mode samples with both $\gamma$=0 and free to vary are reported in the first four lines of Table~\ref{tab:results}. We note that we identified this first set of fits as 'PhotPar' to distinguish it from a different fitting procedure that is described below. As an example of the results of this analysis, Fig.~\ref{fig:parallaxComparison} shows the excellent correlation between the EDR3 and the photometric parallaxes (case with $\gamma$ free to vary).
To have robust uncertainties on the coefficients $\alpha$, $\beta$, and $\gamma$, we adopted a  bootstrap procedure, that is the fit to the data of Eq.~\ref{eq:chi} is repeated 1000 times. For each bootstrap, we obtained a value of $\alpha$, $\beta$, and $\gamma$ and their standard deviations were obtained from the resulting distributions. A detailed description of this procedure can be found in \citet{Ripepi2019}.

Column 7 of Table~\ref{tab:results} provides the reduced $\chi^2$ values obtained from our procedure. Cases 1--2 and 3--4 show the results for $\gamma$=0 and free to vary, respectively. The reduced $\chi^2$ values in absence of a metallicity term are significantly larger than the other cases. In particular, the lowest $\chi^2$ value was obtained for both the F and F+1O sample and $\gamma$ free to vary, that is cases 3 and 4 of Table~\ref{tab:results}. This last case was retained as our best solution due the larger adopted sample.
We also note that the reduced $\chi^2$ value is not close to the expected unity value, indicating that in spite of our thorough treatment of the errors, we still underestimate them. The underestimation can be both in the EDR3 parallax errors and in the photometric parallax uncertainties. For example, if we increase the intrinsic dispersion of the $PW$ in the \gaia\ bands by 50\%, the reduced $\chi^2$ would approach unity. 

To check these results, we adopted a different method to derive the PWZ relation, using the astrometric-based luminosity \citep[ABL][]{Feast1997,Arenou1999}:

\begin{equation}
{\rm ABL}=10^{0.2 W}=10^{0.2(\alpha+\beta\log P +\gamma[Fe/H])}=\varpi10^{0.2w-2}
\label{eqABLZ}
,\end{equation}

\noindent
where, as above, $W$ and $w$ are the absolute and apparent Wesenheit magnitudes and $\varpi$ is the parallax.
We adopted a different fitting procedure with respect to previous calculation, using the nonlinear least square ({\tt nls}) routine included in the {\tt R}
  package\footnote{http://www.R-project.org/}. 
The procedure involves a weighted fitting and a bootstrap method exactly as described above to measure robust errors on the parameters of the fit.  The results obtained with the ABL fitting to the data for the cases with and without a metallicity term and for F and F+1O mode DCEPs  are shown in the last four rows of Table~\ref{tab:results} and identified with the label 'ABL' in the last column of the table.

Comparing the results from the PhotPar and ABL methods, we obtained very similar coefficients for the PWZ relations. The only remarkable difference consists in the smaller $\gamma$ values obtained with the ABL method, but they agree with those of PhotPar within 1$\sigma$. For example, cases 4 and 8 of Table~\ref{tab:results} do indeed provide distances that agree with each other, on average, within $\sim$1\%.

We also note that the ABL method provides larger $\chi^2$ values than the PhotPar case; this is likely the result of a different way of using weights in the minimisation procedure in {\tt R}. However, also for the ABL method, the minimum $\chi^2$ values were obtained for the sample F+1O with the metallicity term included in the calculation (i.e. case 8 in Table~\ref{tab:results}).


We can now compare our $PWZ$ relations with the only previous evaluation available in the literature, by \cite{Ripepi2019}. Using a sample of 261 F DCEPs with DR2 parallaxes and metallicity from the literature, these authors  found:  $W=(-5.996\pm0.082)-(3.134\pm0.095)(\log P-1.0)-(0.237\pm0.199){\rm [Fe/H]}$. The agreement with our F solutions is remarkably good regarding the slope and the intercept, while the metallicity term is smaller by more than $\sim 1.5 \sigma$ with respect to the present work. This occurrence can be explained with both the lower precision of the DR2 parallaxes and the poorer sample of DCEPs adopted in that previous work, indeed the metallicity term in \citet{Ripepi2019} was barely significant at $1\sigma$.  On the other hand, the obtained metallicity dependence seems to be larger than  expectations based on recent non-linear convective pulsation models (De Somma et al. in preparation) that predict a significantly smaller metallicity effect (not larger than 0.2 mag/dex) in period-luminosity-colour (PLC) and PW relations, independently of the filter selection, than in optical PL relations \citep[see e.g.][and references therein]{Caputo2000,Fiorentino2002,Marconi2005}.

As a final note on the size of the metallicity term found in this work, we recall that according to R21, this quantity depends on the adopted global correction to the parallax ZPO. Adopting a larger global correction means reducing the size of the metallicity term. In this respect, it is important, especially for future \gaia\ releases, to have an independent and accurate measure of the parallax ZPO offsets.


   \begin{figure*}
   \hbox{
   \includegraphics[angle=0,width=9.0cm]{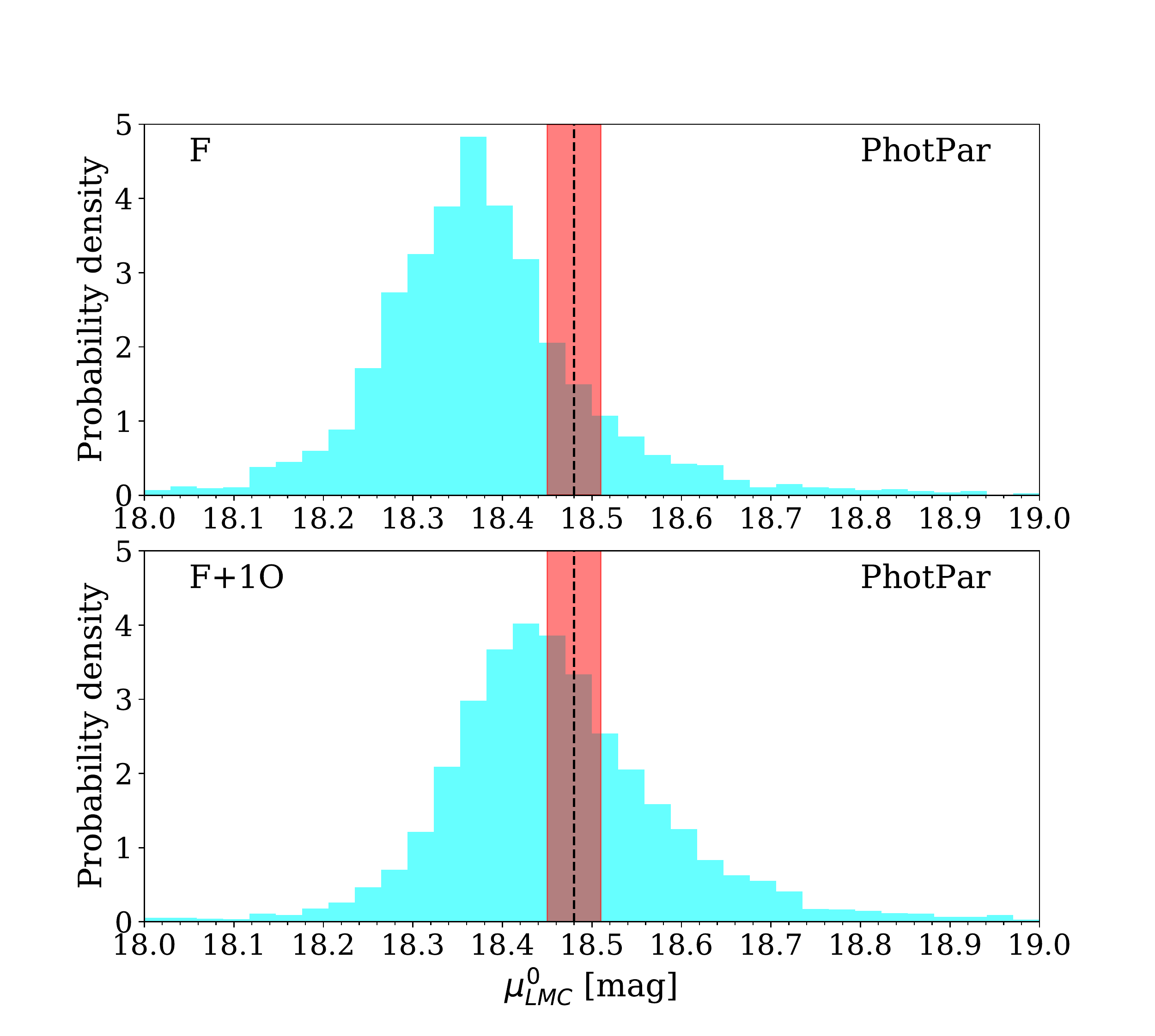}
   \includegraphics[angle=0,width=9.0cm]{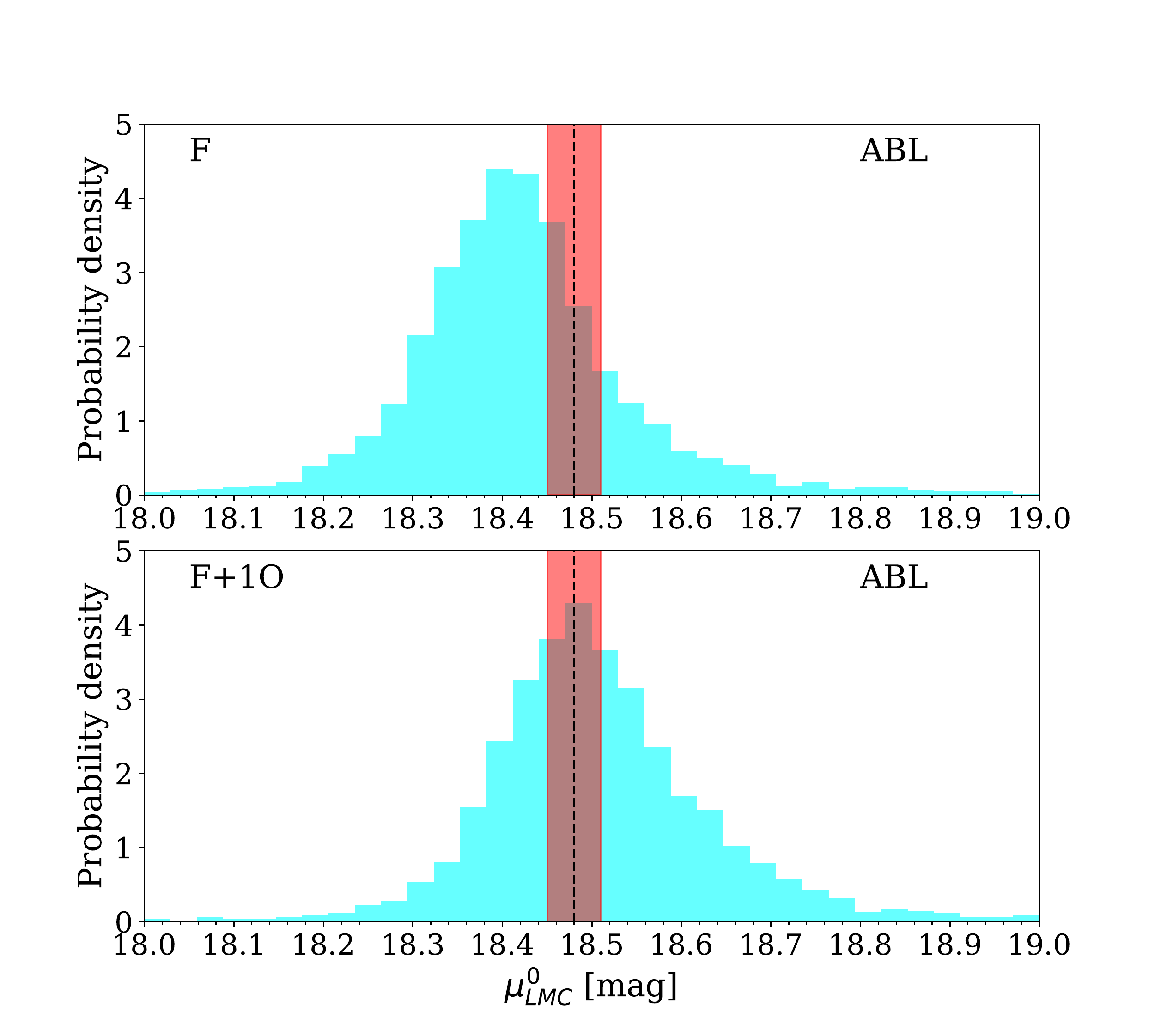}
   }
   \caption{Distribution of the dereddened DM of the LMC calculated by using the PWZ relations listed in Table~\ref{tab:results} (light blue histograms). The left panels show the results obtained with the PhotPar method for the F (top) and F+1O (bottom) mode DCEPs, respectively. The right panels show the same results, but for the ABL method. The red band displays the uncertainty region around the geometric distance of the LMC by \citet[][dashed line]{Pietrzynski2019}.}
              \label{fig:testPLZ}
    \end{figure*}

   \begin{figure}
   \centering
   \includegraphics[angle=0,width=9.0cm]{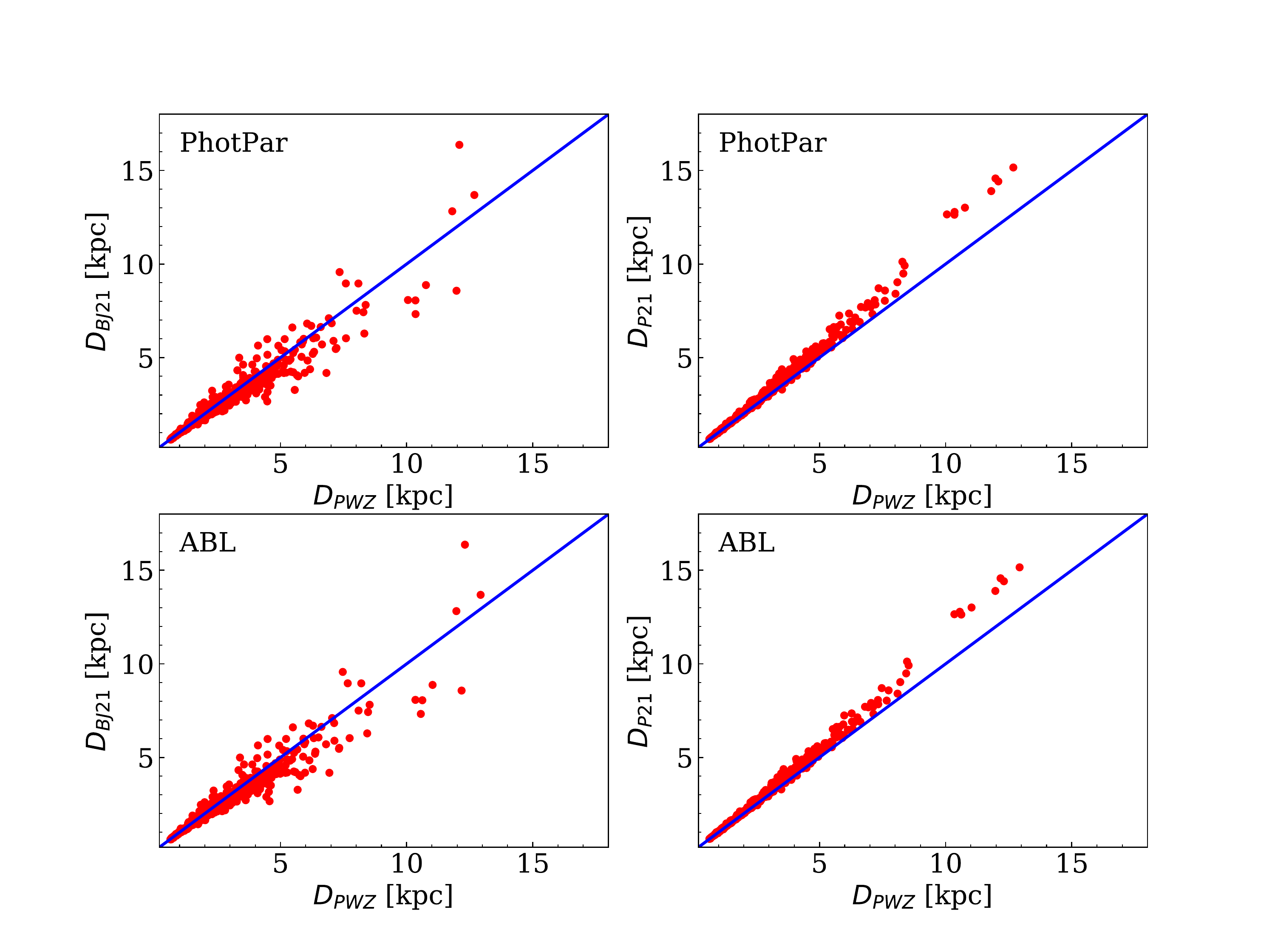}
   \caption{Distance comparison between this work and selected literature results. The left panels show the comparison between distances obtained in this work ($D_{PWZ}$) and those published by \citet{Bailer2021} ($D_{BJ21}$). The right panels are the same as the left ones, but they show the comparison with the distances by \citet{Poggio2021} ($D_{P21}$). The top and bottom panels show the comparisons obtained with the PhotPar and ABL methods, respectively.}
              \label{fig:distanceComparison}%
    \end{figure}

  \begin{figure*}
   \centering
   \hbox{
\includegraphics[angle=0,height=7.3cm]{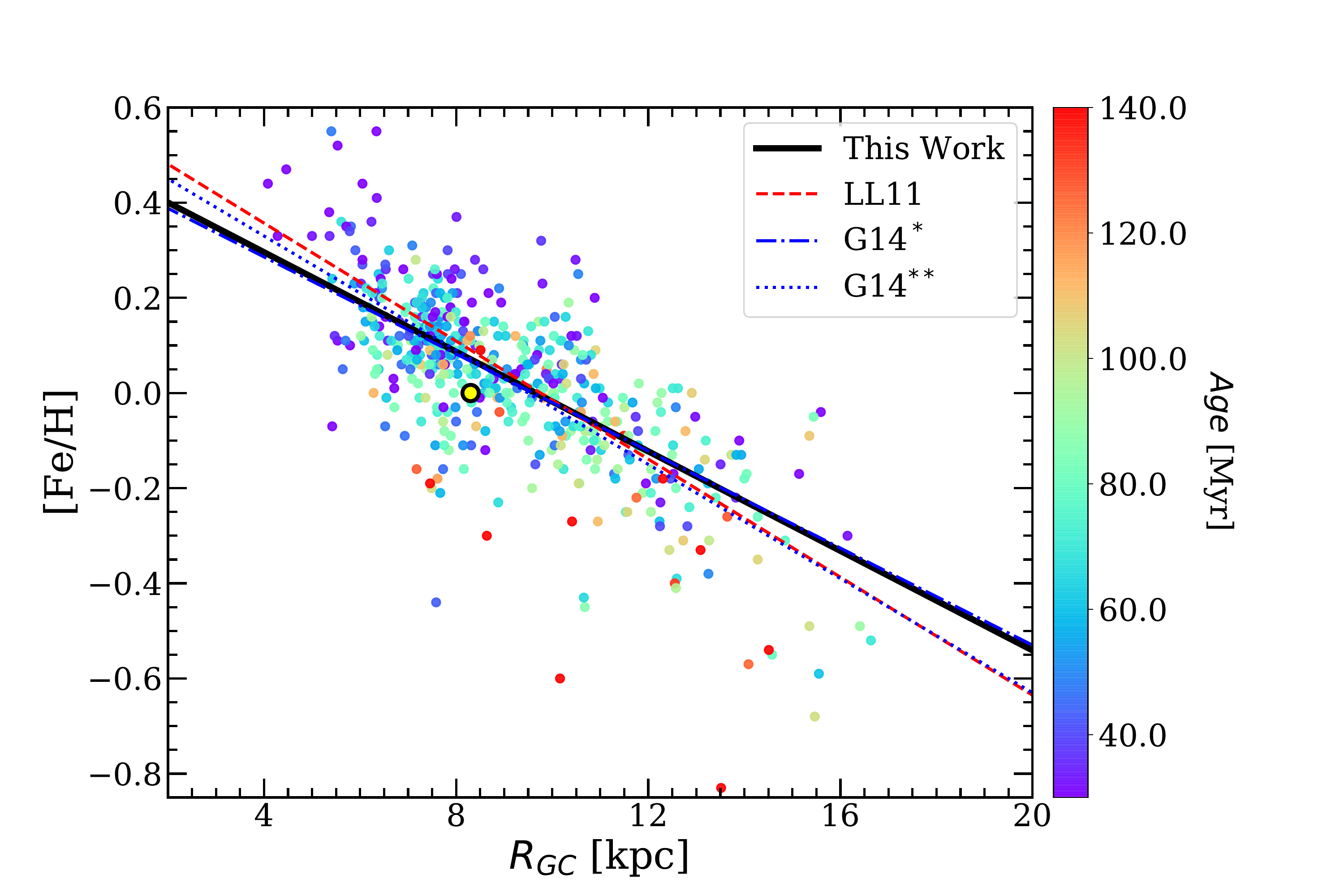}
\includegraphics[angle=0,height=7.3cm]{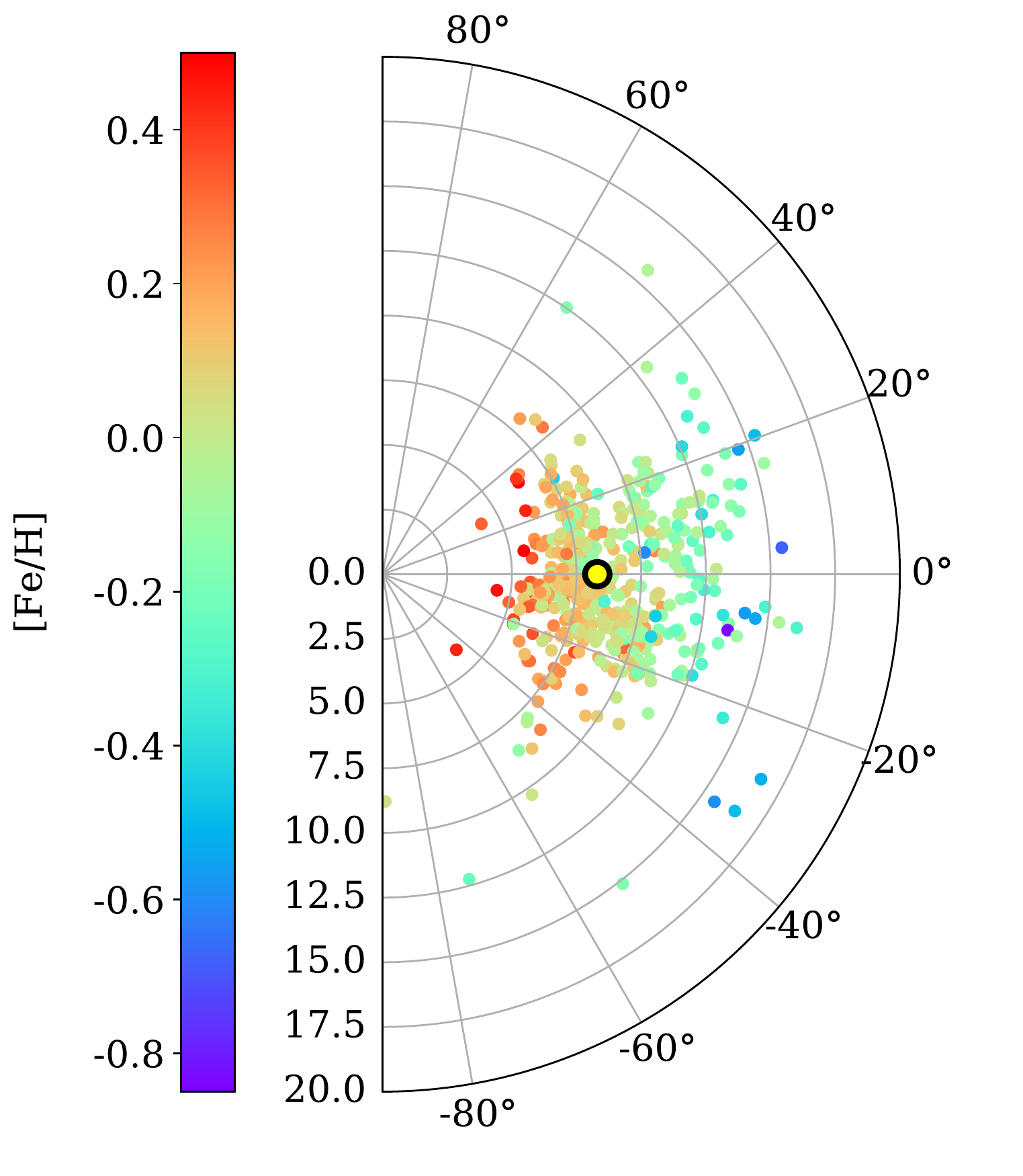}}
   \caption{Metallicity gradient in the Galactic disc inferred from the DCEPs analysed in this work. Left panel: Filled circles represent the sample of DCEPs adopted in this work which  are colour-coded according to their age. The thick black line shows the linear regression to the data obtained in this work (the line is representative of both Eq.~\ref{eq:gradPhotPar} and ~\ref{eq:gradABL}, which are indistinguishable in this diagram).  Selected literature results are shown for comparison: LL11=\citet{Luck2011}; G14$^*$= UVES and FEROS sample from \citet{Genovali2014}; G14$^*$= entire sample of \citet{Genovali2014}. Right panel: Polar representation of the DCEPs considered in this work. Dots are colour-coded according to the DCEPs' metallicity. In both panels, a yellow and black circle represents the Sun.}
              
              \label{fig:metGradient}%
    \end{figure*}

\section{Discussion}

We tested the reliability of the $PWZ$ relation derived in this work in several different ways.

\subsection{Distance to the Large Magellanic Cloud}

As a first test, we applied our $PWZ$ relations to the DCEPs in the LMC, derived the LMC  mean dereddened distance modulus $\mu^0_{LMC}$, and compared it to the geometric estimate from eclipsing binaries by \citet{Pietrzynski2019}. The latter,  $\mu_{LMC}=18.48\pm0.03$ mag (including systematic errors), is considered one of the most accurate estimates in the literature to date.

To this aim, we considered a sample of about 4500 DCEPs in the LMC with periods published by the OGLE IV survey \citep[The Optical Gravitational Lensing Survey IV,][]{Udalski2018} and cross-matched their positions with the EDR3 catalogue to obtain the $G,G_{BP},G_{RP}$ magnitudes needed to calculate the apparent Wesenheit magnitudes, $w$.

Then, we calculated the absolute Wesenheit magnitude $W$ for each LMC DCEP, adopting the coefficients of the $PW$/$PWZ$ relations reported in Table~\ref{tab:results}, using the OGLE IV periods and assuming ${\rm [Fe/H]}_{LMC}=-0.407\pm0.003$ dex (dispersion $\sigma = 0.076\pm0.003$ dex), according to the recent evaluation by \citet[][]{Romaniello2021}.
 From these $W$ values, we calculated individual distance moduli for each LMC DCEP
 as $\mu_{LMC}$=$w-W$, obtaining a distribution whose median gives the estimate of $\mu_{LMC}$. The error on this value was calculated by performing a set of 1000 Monte Carlo simulations. Specifically, we varied the $PW$/$PWZ$ relations generating new $\alpha$ $\beta$ and $\gamma$ coefficients extracted from normal distributions centred on the fitted values of Table~\ref{tab:results} and with standard deviations given by the respective errors. For every experiment, we re-calculated the LMC median distance. The provided final error was estimated by taking the robust standard deviation of the obtained sample of 1000 mean distances. In this process, we neglected the metallicity dispersion of LMC DCEPs, as we verified that it is too small ($\sim$0.07 dex) to affect the distances.  
 
Final values of $\mu_{LMC}$ and related errors are listed in Table~\ref{tab:results}, along with the number of LMC DCEPs adopted for the calculation. Starting from the cases with a null metallicity term ($\gamma=0$, cases 1--2, and 5--6 in Table~\ref{tab:results}), it can be seen that in all the cases, the $\mu_{LMC}$ values are larger by $\sim 6 \sigma$ than the \citet{Pietrzynski2019} value, an occurrence that confirms the importance of introducing a metallicity term in the $PW$ relation. 
Now considering the values of $\mu_{LMC}$  obtained for F and F+1O samples and $\gamma$ free to vary, we see that they are in agreement with each other within 1$\sigma$, the difference being explained by the larger metallicity term found for the F case. 
The comparison between our LMC distances and the geometric estimate by \citet{Pietrzynski2019} is shown in Fig.~\ref{fig:testPLZ}.
All the four cases agree well within 1$\sigma$ with the geometric estimate, even if the better match between the LMC distance distribution and the \citet{Pietrzynski2019} value is obtained for case 8 of Table ~\ref{tab:results}, that is to say with the F+1O $PWZ$ relation  derived with the ABL method. We consider this case as our best $PWZ$ relation.

\subsection{Distances in the MW}

As a second test, we compared the distances derived from our $PWZ$ relation  with the distances 
derived from a Bayesian treatment of the parallaxes by 
\citet{Bailer2021}\footnote{They published two different distance estimates, one purely geometric, based on the astrometry, and the other 'photogeometric' distance, based on both photometry and astrometry.  Here we used the purely geometric one, but adopting the other distance provides the same results.} 
and with the distances derived in our previous work \citep[][]{Poggio2021}, which are based on a $PW$ relation in the \gaia\ bands 
calibrated on a larger DCEP sample but 
without 
including a metallicity term. The result of this comparison is shown in Fig.~\ref{fig:distanceComparison}. First, we note that there are no detectable differences between the use of the PhotPar or ABL method. In both cases, our distances are in good agreement (better than $\pm$10\%) with those by \citet{Bailer2021} in the first 2.5 kpc from the Sun. Beyond this value, our distances tend to be increasingly larger on average, with a high scatter at large distances due to the progressive decrease in the accuracy of the \gaia\ parallaxes. The comparison with \citet[][]{Poggio2021} shows that our distances are smaller, but the difference is contained within 10\% for 85\% of the DCEPs even if the discrepancy becomes more significant beyond 5-6 kpc. The reason for this difference mainly resides in the inclusion of the metallicity term as well as in the different choices in terms of ZPO of the \gaia\ parallaxes. Indeed, while in \citet{Poggio2021}, only the \citet{Lindegren2021} individual parallax ZPOs  we used, here, in addition to those, we also applied the global parallax ZPO correction by \citet{Riess2021}. 

\subsection{Metallicity gradient of the MW disc}

As a further test, we computed the metallicity gradient of the disc based on the 499 DCEPs used in the present work and compared it with  literature values. As a first step, we determined the Galactocentric radius of each DCEP in our sample. To this aim, we adopted the same procedure as in section 3.2.2 of \citet{Ripepi2019}, using $D_0$=8.0$\pm$0.3 kpc for the Galactocentric distance to the Sun \citep[][]{Camarillo2018}.  

The variation of ${\rm [Fe/H]}$ with  Galactocentric radius,  $R_{GC}$,  is shown in Fig.~\ref{fig:metGradient} (left panel). The right panel of the figure instead shows the variation of ${\rm [Fe/H]}$ in polar coordinates. We carried out a linear regression to the data using the {\tt python LtsFit} package \citep{Cappellari2013}, which allows one to use weights on both variables 
as well as an extremely robust outlier removal.
To be conservative, we adopted a 3$\sigma$ clipping procedure, which led us to exclude ten objects.
The metallicity gradient derived with this procedure is based on 489 DCEPs and is described by the following linear relations for the PhotPar and ABL methods: 

\begin{eqnarray}
    {\rm [Fe/H]}=(-0.0523\pm0.0024) R_{GC}+(0.505\pm0.022) \label{eq:gradPhotPar}\\
    {\rm [Fe/H]}=(-0.0527\pm0.0022) R_{GC}+(0.511\pm0.022) \label{eq:gradABL}
\end{eqnarray}

\noindent 
with rms=0.11 dex in both cases. The two solutions are statistically indistinguishable, given the slightly smaller error on the gradient, and we consider Eq.~\ref{eq:gradABL} as our best value. A comparison between our result  based on DCEPs and the gradients in the recent literature obtained with a similar technique is shown in Table~\ref{tab:comparison} and Fig.~\ref{fig:metGradient} (left panel).

\begin{table}
\caption{Comparison between the Galactic metallicity gradient derived in this work and the literature values. 
The functional form is ${\rm [Fe/H]}=a \times R_{GC}+b$. The values of the slope $a$ and intercept $b$ are listed in column 1 and 2, respectively. Column 3 reports the number of sources used for the fit, while column 4 provides the literature source.}
\label{tab:comparison} 
\footnotesize\setlength{\tabcolsep}{3pt}
\centering          
\begin{tabular}{cccc} 
\hline\hline             
$a$ & $b$ & n.DCEPs & source \\
(dex/kpc) & (dex)  &  & \\ 
\hline
 $-0.062 \pm 0.002$ & 0.605$\pm$0.021 & 313   & \citet{Luck2011} \\ 
 $-0.051\pm0.003$ & $0.49\pm0.03$ & 128 & \citet{Genovali2014}$^*$ \\
 $-0.060\pm0.002$ & $0.57\pm0.02$ & 450 & \citet{Genovali2014}$^{**}$ \\
 $-0.051 \pm 0.002$ &  & 411   & \citet{Luck2018} \\ 
 $-0.0523\pm0.0024$ & $0.505\pm 0.022$ & 489 & This work$^{\dagger}$ \\
 $-0.0527\pm0.0022$ & $0.511\pm 0.022$ & 489 & This work$^{\dagger \dagger}$ \\
 \hline                                   
\end{tabular}
\tablefoot{$^*$ values obtained using UVES and FEROS spectroscopy only (see text)\\
$^{**}$ values obtained with the whole sample \\ 
$^{\dagger}$ values obtained with the PhotPar method\\ 
$^{\dagger \dagger}$ values obtained with the ABL method
}
\end{table}

Our 
result is in good agreement with the first evaluation by \citet{Genovali2014}\footnote{This result is based on a sample of 128 DCEPs having metallicities measured with UVES (Ultraviolet and Visual Echelle Spectrograph) and FEROS (The Fiber-fed Extended Range Optical Spectrograph).} and with the recent work by \citet{Luck2018}. Instead, we disagree with the second evaluation by \citet{Genovali2014} (obtained adding literature data for 322 DCEPs to the previous dataset) and with that by \citet{Luck2011}. 
All the aforementioned works find an  intrinsic scatter of  the order of 0.10-0.12 dex, which is in agreement with our result.
The right panel of Fig.~\ref{fig:metGradient} shows the variation of  ${\rm [Fe/H]}$, not only as a function of the Galactocentric distance, but also depending on the direction. It can be seen that the metallicity gradient appears to be constant in all  directions, again in agreement with \citet{Luck2018}. 

Before concluding, we note that the left panel of Fig.~\ref{fig:metGradient} also reports the age of the DCEPs analysed in this work, where ages were calculated using the period-age-metallicity relation by \citet{Desomma2021}. In particular, we show the ages obtained using their relation A for F-mode pulsators (calculated using models without overshooting, see their Table 9). However, we verified that the use of the relation B (models with overshooting) does not alter the general trend of the ages.
The figure reveals that the most metal-rich objects with ${\rm [Fe/H]}>0.3$ dex (closer to Galactic centre) all have ages smaller than $\sim$50 Myr. In general, the DCEPs younger than $\sim$80 Myr tend to stay above the mean gradient line, while the reversed behaviour can be seen for the DCEPs older than $\sim$120 Myr, which are therefore older than the more metal-rich ones located at the same Galactocentric distance. It is difficult to explain this occurrence with the  age-metallicity relation, as the age difference between the DCEPs is too short to justify the observed metallicity difference ($\Delta {\rm [Fe/H]} \sim 0.2-0.4$ dex). A possible explanation is the mixing of DCEPs coming from different regions of the disc. However, a detailed investigation of this point is beyond the scope of this work.

\section{Conclusions}

In this paper we have investigated the metallicity dependence of the Galactic DCEP PW relation in the \gaia\ bands.  
In particular, we used a sample of 435 DCEPs with metallicity measurements from high-resolution spectroscopy, in conjunction with \gaia\ parallaxes and photometry from EDR3 to calibrate a $PWZ$ relation in the \gaia\ bands. We adopted two different fitting procedures to calculate the coeffcient of the $PWZ$ relations, providing robust uncertainties by means of the bootstrap technique. We find a significant metallicity term, of the order of $-$0.5 mag/dex, which is larger than what was measured in the NIR bands by different authors \citep[e.g.][]{Breuval2021,Riess2021,Ripepi2021}. Our best $PWZ$ relation is $W=(-5.988\pm0.018)-(3.176\pm0.044)(\log P-1.0)-(0.520\pm0.090){\rm [Fe/H]}$.

We validated our $PWZ$ relations by using the distance to  the LMC as a benchmark, finding very good agreement with the geometric distance provided by \citet{Pietrzynski2019} based on eclipsing binaries. On the contrary, the $PW$ relations without a metallicity term provide LMC distances larger by $\sim 6 \sigma$ with respect to this value. 

As an additional test, we used 489 DCEPs in  our sample to evaluate the metallicity gradient in the young Galactic disc, finding values of $-$0.0523$\pm$0.0024 dex/kpc or  $-$0.0527$\pm$0.0022 dex/kpc (PhotPar and ABL methods, respectively), which are in very good agreement with previous results.

The $PWZ$ relations presented in this work will be crucial to fully exploit the results of the forthcoming \gaia\ DR3 as they will allow us to use DCEPs to study, with unprecedented detail, the structure and dynamics of the Galactic spiral arms, where most DCEPs reside, up to the farthest regions, where distances from parallaxes will be hampered by large errors or will not be available at all.

\begin{acknowledgements}

We wish to thank our anonymous Referee whose pertinent and constructive comments helped us to improve the manuscript.
This work has made use of data from the European Space Agency (ESA) mission
{\it Gaia} (\url{https://www.cosmos.esa.int/gaia}), processed by the {\it Gaia}
Data Processing and Analysis Consortium (DPAC,
\url{https://www.cosmos.esa.int/web/gaia/dpac/consortium}).
Funding for the DPAC has been provided by national institutions, in particular the institutions participating in the {\it Gaia} Multilateral Agreement.
In particular, the Italian participation
in DPAC has been supported by Istituto Nazionale di Astrofisica
(INAF) and the Agenzia Spaziale Italiana (ASI) through grants I/037/08/0,
I/058/10/0, 2014-025-R.0, and 2014-025-R.1.2015 to INAF (PI M.G. Lattanzi).
V.R., M.M. and G.C. acknowledge partial support from the project 'MITiC: MIning The Cosmos Big Data and Innovative Italian Technology for Frontier Astrophysics and Cosmology'  (PI B. Garilli).
This research has made use of the SIMBAD database,
operated at CDS, Strasbourg, France.
\end{acknowledgements}

%
%

\bibliographystyle{aa} 
\bibliography{myBib} 

\end{document}